\documentclass[12pt]{article}
\usepackage{amsmath,amssymb}
\usepackage{hyperref}

\newcommand{\bra}[1]{\langle #1\vert}
\newcommand{\ket}[1]{\vert #1\rangle}

\newcommand{\eg}{e.\,g.\ }
\newcommand{\ie}{i.\,e.\ }

\newcommand{\be}{\begin{equation}}
\newcommand{\ee}{\end{equation}}
\newcommand{\ra}{\rightarrow}

\renewcommand{\Im}{{\rm Im}}
\renewcommand{\epsilon}{\varepsilon}
\renewcommand{\exp}[1]{\langle #1\rangle}

\title{Leggett--Garg Inequalities, Pilot Waves and Contextuality}
\author{Guido Bacciagaluppi\thanks{Department of Philosophy, University of Aberdeen; and UMR 8590 IHPST-Institut d'Histoire et de Philosophie des Sciences et des Techniques, Universit\'{e} Paris 1 Panth\'{e}on-Sorbonne, CNRS, ENS. Address for correspondence: Department of Philosophy, University of Aberdeen, Old Brewery, High Street, Aberdeen AB24 3UB, Scotland (e-mail: g.bacciagaluppi@abdn.ac.uk).}}

\date{16 November 2014}

\begin{document}

\maketitle

\begin{abstract}
In this paper we first analyse Leggett and Garg's argument to the effect that macroscopic realism contradicts quantum mechanics. After making explicit  all the assumptions in Leggett and Garg's reasoning, we argue against the plausibility of their auxiliary assumption of non-invasive measurability, using Bell's construction of stochastic pilot-wave theories as a counterexample. Violations of the Leggett--Garg inequality thus do not provide a good argument against macrorealism {\em per se}. We then apply Dzhafarov and Kujala's analysis of contextuality in the presence of signalling to the case of the Leggett--Garg inequalities, with rather surprising results. An analogy with pilot-wave theory again helps to clarify the situation.

\

\
\end{abstract}

\section{Introduction}\label{intro}
Leggett and Garg (1985) establish a contradiction between quantum mechanics and what they call the assumptions of `macroscopic realism' and `non-invasive measurability on the macroscopic scale', using the example of macroscopic quantum tunnelling in SQUIDs previously discussed by Chakravarty and Leggett (1984). 

In this paper, we shall first of all,  in Section~\ref{inequalities}, review Leggett and Garg's argument, making more explicit the assumptions needed to establish their result. In particular, we shall distinguish two components in the assumption of non-invasive measurability (see also Leggett (2002)): (a) that one can always perform measurements that reveal the pre-existing value of the macroscopic flux (we shall call this `faithful measurability'), and (b) that one can always perform measurements of the macroscopic magnetic flux that do not alter the subsequent dynamics of the system (for this we shall use Leggett and Garg's own term of `non-invasive measurability'). The contradiction is between quantum mechanics on the one hand and the assumptions of macroscopic realism and simultaneously faithful and non-invasive measurability on the other hand. 

While Leggett and Garg believe that macroscopic realism plausibly implies (simultaneously faithful and) non-invasive measurability, so that violation of their inequality in fact rules out macroscopic realism, we point out in Section~\ref{invasive} that models of the quantum mechanical predictions satisfying both macroscopic realism and faithful measurability are easy to construct, along the lines of Bell's (1986) stochastic pilot-wave theories, but that these naturally violate non-invasive measurability. Thus, while violation of the Leggett--Garg inequality may indeed be a signature of quantum behaviour on a macroscopic scale, it is not sufficient to rule out macroscopic realism as such.

Section~\ref{context} then analyses the violation of the Leggett--Garg inequality in terms of the discussion by Dzhafarov and Kujala (2014a,b) of contextuality in the presence of temporal signalling (i.e.\ for the case in which performing a measurement affects the distribution of results of a later measurement). Surprisingly, in the case considered by Leggett and Garg, namely with a pure initial state, the violation of the Leggett--Garg inequality turns out not to be sufficient for contextuality in the sense of Dzhafarov and Kujala, because of the amount of signalling present. Instead, if the initial state is maximally mixed, the situation is reversed: no temporal signalling is possible and the violation is entirely described in terms of contextuality. 

Our final Section~\ref{analogy} again invokes pilot-wave theory to provide a helpful analogy for explaining the discrepancy between the pure and mixed cases. The disanalogy, however, remains in the case of `Schr\"{o}dinger's SQUID'.\footnote{For two other recent and very penetrating criticisms of Leggett and Garg (both partially overlapping with the present discussion), see Kofler and Brukner (2013) and Maroney and Timpson (in preparation).}

\section{The Leggett--Garg inequalities}\label{inequalities}
The example of macroscopic quantum tunnelling used by Leggett and Garg (1985) is essentially the familiar one of quantum tunnelling in a symmetric double potential well: for finite barrier height the ground state energy is split into $E\pm\frac{\Delta E}{2}$, the degeneracy between the symmetric and anti-symmetric states $\ket{\psi_S}$ and $\ket{\psi_A}$ is lifted, and the system oscillates between the states $\ket{\psi_R}:=\frac{1}{\sqrt 2}(\ket{\psi_A}-\ket{\psi_S})$ and  $\ket{\psi_L}:=\frac{1}{\sqrt 2}(\ket{\psi_A}+\ket{\psi_S})$ localised in the two wells:
  \begin{equation}
    \ket{\psi(t)}=\cos(\tfrac{\Delta E}{2\hbar}(t-t_0))\ket{\psi_R}-i\sin(\tfrac{\Delta E}{2\hbar}(t-t_0))\ket{\psi_L} \ .
  \label{a}
  \end{equation}
Setting $Q:=\ket{\psi_R}\bra{\psi_R}-\ket{\psi_L}\bra{\psi_L}$, from elementary trigonometric identities one can calculate the following quantum mechanical expectation values:
  \be
    \exp{Q(t)}=\cos(\tfrac{\Delta E}{\hbar}(t-t_0)) 
  \label{b}
  \ee
(where the argument of $Q$ refers to the time of measurement), and more generally
  \be
    \exp{Q(t_j)Q(t_i)}=\cos(\tfrac{\Delta E}{\hbar}(t_j-t_i))
  \label{c}
  \ee
(understood as the expectation value for the results of two measurements in succession, with intervening collapse of the wave function\footnote{In the notation we introduce later, we shall be writing $\exp{Q(t_i)}=\exp{Q^i_i}$ and $\exp{Q(t_j)Q(t_i)}=\exp{Q^{ij}_iQ^{ij}_j}$.}). Note for future reference (we shall use this in footnote 6 below and especially in Section~\ref{context}) that (\ref{c}) is independent of $t_0$, so it remains the same if the initial state at $t=0$ is an arbitrary state of the form (\ref{a}), e.g.\ $\ket{\psi_R}$ or $\ket{\psi_L}$, and consequently also if it is an arbitrary mixture of such states.

Given that the system is a macroscopic magnetic flux, environmental effects are unavoidable, and coherent behaviour such as the above is generally suppressed. Chakravarty and Leggett (1984), however, show that in a certain regime and on a certain time scale the oscillations are merely underdamped, so that qualitatively the system behaves like the familiar case, and one has indeed {\em macroscopic\/} quantum tunnelling.

Leggett and Garg now use this as a test case for (or rather against) common intuitions that a macroscopic system should always be in a macroscopically well-defined state, and thus (if $Q$ specifically refers to the macroscopic magnetic flux in the SQUID) that $Q$ should always {\em have\/} the value $+1$ or $-1$, irrespective of measurement. A direct confirmation that the system is in a non-trivial superposition (\ref{a}) through a measurement of an observable $P(t):=\ket{\psi(t)}\bra{\psi(t)}$ is not in itself feasible, but Leggett and Garg point out that, even if one is able to measure only the observable $Q$,  one obtains a contradiction with the predictions of quantum mechanics by considering sequential measurements at times $t_1, t_2, t_3,\ldots$ (which, because of the non-trivial dynamics (\ref{a}), are effectively sequential measurements of non-commuting observables). More precisely, they argue that a contradiction follows under the additional assumption (to be spelled out below) that $Q$ can be measured non-invasively --- which they consider plausible given macroscopic realism. 

Here are the details (as we see them). As mentioned, Leggett and Garg assume what they call macroscopic realism, by which they explicitly mean that the macroscopic quantity $Q$ always has a definite value. Implicitly, their assumption of macroscopic realism further includes the existence of a (generally stochastic) evolution for $Q$ so that all the probability distributions
  \begin{equation}
    \rho(Q_1,Q_2,Q_3,\ldots)
  \label{1}
  \end{equation}
are well-defined, where $Q_i$ is the random variable $Q$ at time $t_i$. It then follows that various Bell inequalities are satisfied for the pairwise marginals derived from (\ref{1}), for instance
  \be
    1+\exp{Q_1Q_2}+\exp{Q_2Q_3}+\exp{Q_1Q_3}\geq 0 \ ,
  \label{d}
  \ee
which we shall refer to as `the Leggett--Garg inequality', and which is one-half of the inequality by Suppes and Zanotti (1981),\footnote{Both inequalities (\ref{d}) and (\ref{SZ}) are a special case of the CHSH inequalities for the case in which
one pair of observables is equal (so that the expectation of their product is identically 1). However, while Leggett and Garg are interested in (\ref{d}) as a necessary condition for the existence of joint distribution (\ref{1}), the main result in Suppes and Zanotti (1981) was the first published proof (under the restriction of flat marginals) that a Bell-type inequality is a {\em necessary and sufficient\/} condition for the existence of a joint probability distribution. The necessity and sufficiency of the CHSH inequalities was established at the same time by Fine (1982). Thanks to Ehtibar Dzhafarov for correspondence on these points.}
  \begin{multline}
    -1\leq \exp{Q_1Q_2}+\exp{Q_2Q_3}+\exp{Q_1Q_3}\leq \\
    \leq1+2\min\{\exp{Q_1Q_2},\exp{Q_2Q_3},\exp{Q_1Q_3}\} \ .
  \label{SZ}
  \end{multline}

The random variables $Q_j$ in (\ref{1}) refer to the system as it evolves on its own, with no measurements being performed. In order to consider also measurements on the system, let us introduce new random variables $Q^{\ldots ijk\ldots}_j$, etc., re\-presenting the magnetic flux at the time $t_j$ for the case in which measurements are carried out at the times $\ldots t_i,t_j,t_k,\ldots$

Explicitly, Leggett and Garg now make the assumption of {\em non-invasive\/} measurability on the macroscopic scale, which they define as follows: `It is possible, in principle, to determine the state of the system with arbitrarily small perturbation on its subsequent dynamics' (1985, p.~857). If one assumes that there are measurements on the system that do not affect the system's trajectory in any way, then it is clear that the probabilities (\ref{1}) equal the directly observable probabilities
  \begin{equation}
    \rho(Q^{123\ldots}_1,Q^{123\ldots}_2,Q^{123\ldots}_3,\ldots) \ ,
  \end{equation}
and that, in particular, 
  \be
     \rho(Q^{ij}_i,Q^{ij}_j)=\rho(Q_i,Q_j) \ ,
  \ee
so one can test whether (\ref{d}) is violated.

Leggett and Garg consider non-invasive measurability in this sense to be plausible given macroscopic realism, and look at the example of ideal negative-result experiments to substantiate this,\footnote{To the best of my knowledge, negative-result experiments in the context of quantum mechanics were first considered by Schr\"{o}dinger (1934, p.~519), as a criticism of the idea that quantum measurements always involve the exchange of at least one quantum of action.} citing the example of a two-slit experiment in which one puts a (perfect) detector behind one slit only, and the detector does not fire. By extension, one might (ideally) check whether the flux is in the right well by means of an interaction that vanishes everywhere else. If the result is negative, one infers that the flux is in the left well, and by macroscopic realism that it was so also immediately before the experiment, and thus --- so Leggett and Garg --- that the experiment has not altered the dynamics of the system. 

We shall return to the plausibility of non-invasive measurability in Section~\ref{invasive} below. Suffice it to say now that this
argument does not seem to establish the assumption in full, in that it does not seem to establish that the system will continue to evolve according to the original dynamics {\em also in the future}; but it certainly establishes the plausibility of the more modest assumption that it is possible to carry out {\em faithful\/} measurements at the macroscopic scale, in the sense that it is possible to carry out a measurement of $Q_i$ such that the corresponding result indeed equals the pre-existing value of $Q_i$. 

Accordingly, for the purposes of Section~\ref{invasive} it will be useful to distinguish explicitly two components in Leggett and Garg's non-invasive measurability, namely that it is possible to perform a measurement that: (a) does not alter the value of $Q$ that is being measured (we shall call this `faithful measurability'), and (b) does not alter the transition probabilities for the values of $Q$ at later times (for this we shall use Leggett and Garg's original term `non-invasive measurability').\footnote{More recently, Leggett (2002) has also been explicitly including both components in the definition of non-invasive measurability.}

Assumption (a) that measurements of $Q$ are faithful means that, for all sequences of measurements at times $\ldots t_i\leq t_j\leq t_k,\ldots$, the following identity of random variables holds:
  \begin{equation}
    Q^{\ldots ijk\ldots}_j=Q^{\ldots ik\ldots}_j \ ,
  \label{1b}
  \end{equation}
and in particular that 
  \begin{equation}
    Q^i_i=Q_i \ .
  \label{2}
  \end{equation}

Assumption (b) that measurements are non-invasive (in our more specific sense) means that, for all sequences of measurements at times $\dots t_i\leq t_j\leq t_k$ and for all values of the flux at times $t_p\leq t_q\ldots \leq t_r$, if $t_r>t_k$ then  
  \begin{equation}
    \rho(Q^{\ldots ijk}_r|Q^{\ldots ijk}_p,Q^{\ldots ijk}_q,\dots)=
    \rho(Q^{\ldots ij}_r|Q^{\ldots ij}_p,Q^{\ldots ij}_q,\ldots) \ ,
  \label{3} 
  \end{equation} 
or equivalently (since $t_r>t_k\geq t_j\geq t_i\ldots$), if $t_r>t_k$ then 
  \begin{equation}
    \rho(Q^{\ldots ijk}_r|Q^{\ldots ijk}_p,Q^{\ldots ijk}_q,\dots)=\rho(Q_r|Q_p,Q_q,\ldots) \ .
  \label{3bis} 
  \end{equation} 

It now obviously follows again that, given macroscopic realism, if one assumes that measurements of $Q$ are both faithful and non-invasive, one has
  \begin{equation}
    \rho(Q^{12\ldots n}_1,Q^{12\ldots n}_2,\ldots, Q^{12\ldots n}_n)=\rho(Q_1,Q_2,\ldots, Q_n) 
    \label{same}
  \end{equation}
for all $n$, as one can easily prove explicitly by induction from (\ref{2}) and (\ref{3}) and the assumption that, for all sequences of measurements at times $\ldots t_i\leq t_j\leq t_k$, if $t_k>t_j$ then $Q^{\ldots ijk}_j=Q^{\ldots ij}_j$ (the identity of random variables does not depend on whether any even later measurements are performed). In particular, one has
  \be
     \rho(Q^{ij}_i,Q^{ij}_j)=\rho(Q_i,Q_j) \ .
     \label{all}
  \ee
This is all that is needed to test the Leggett--Garg inequality,\footnote{Note there are two ways one might go about
testing the Leggett--Garg inequality: (i) by measuring the expectation values $\exp{Q^{ij}_iQ^{ij}_j}$ in three sets of experiments, one for each pair $(i,j)$, or (ii) by measuring the expectation value $\exp{Q^{13}_1Q^{13}_3}$ in one set of experiments, and the expectation values $\exp{Q^{123}_1Q^{123}_2}$ and $\exp{Q^{123}_2Q^{123}_3}$ in a second set of experiments. The two procedures are equivalent as tests of (\ref{d}) because  $\exp{Q^{123}_1Q^{123}_2}=\exp{Q^{12}_1Q^{12}_2}$ (no effect of later measurements) and because $\exp{Q^{123}_2Q^{123}_3}=\exp{Q^{23}_2Q^{23}_3}$ both according to the hypothesis of macroscopic realism (and faithful and non-invasive measurability) and according to quantum mechanics (because measuring $Q$ at $t_1$ leaves the system in the equal-weight mixture of $\ket{\psi_R}$ and $\ket{\psi_L}$, and hence the quantum mechanical expectation value $\exp{Q_2Q_3}$ is given by (\ref{c}) whether or not the measurement at $t_1$ is carried out). We shall return to these two possible scenarios in footnote 17 at the end of Section~\ref{context}.} which becomes
  \be
    1+\exp{Q^{12}_1Q^{12}_2}+\exp{Q^{23}_2Q^{23}_3}+\exp{Q^{13}_1Q^{13}_3}\geq 0 \ .
  \label{d'}
  \ee
However, quantum mechanics predicts that the observable probabilities in fact violate this inequality. 

Indeed, substituting (\ref{c}) into (\ref{d'}) yields
  \be
    1+\cos(\tfrac{\Delta E}{\hbar}(t_2-t_1))+\cos(\tfrac{\Delta E}{\hbar}(t_3-t_2))+\cos(\tfrac{\Delta E}{\hbar}(t_3-t_1))\geq 0 \ ,
  \label{g}
  \ee
and it is easy to show that (\ref{g}) is violated, and maximally so, if one chooses\footnote{Writing (\ref{h}) as  $1+\cos(\alpha)+\cos(\beta)+\cos(\alpha+\beta)$, stationarity with respect to both $\alpha$ and $\beta$ implies $\sin(\alpha)=\sin(\beta)$, \ie $\alpha=\pi-\beta\mod(2\pi)$ or $\alpha=\beta\mod(2\pi)$. In the first case, (\ref{g}) is saturated and thus never violated; in the second case, a simple minimisation yields maximal violation for $\alpha=\frac{2\pi}{3}$.} 
  \be
    \tfrac{\Delta E}{\hbar}(t_3-t_2)=\tfrac{\Delta E}{\hbar}(t_2-t_1)=\tfrac{2\pi}{3} \ ,
  \label{h}
  \ee
for which all three cosine terms  equal $-\frac{1}{2}$. The corrections to (\ref{c}) in the Chakravarty--Leggett model do not essentially affect this result, apart from an adjustment in the effective oscillation frequency, and, indeed, also Leggett and Garg choose the value $\frac{2\pi}{3}$ in order to violate (\ref{d}).\footnote{We cannot reproduce Leggett and Garg's equation (3), but it seems inessential to deriving their results.}

Such a violation correctly implies that at least one of macrorealism, faithful measurability or non-invasive measurability fail (with the last two apparently being plausible consequences of macrorealism).

\section{Pilot-waves and invasive measurements}\label{invasive}
From the above, it would seem that the natural conclusion from the violation of the Leggett--Garg inequality is that macroscopic realism fails. We now wish to argue that the weakest of the three assumptions needed to establish Leggett and Garg's result is that of non-invasive measurability. Indeed, we shall point out that it is a familiar feature of a rather well-known family of theories, satisfying both the assumptions of realism and of faithful measurability, that non-invasive measurability should fail. The best-known example of such a theory is de~Broglie--Bohm pilot-wave theory. In order to see this we shall just need to recall a few basic features of the theory and look at the elementary case of single-slit diffraction (or double-slit diffraction, as in Leggett and Garg's negative-result example). 

As presented by de~Broglie in October 1927 at the fifth Solvay conference (de~Broglie 1928), pilot-wave theory is a novel dynamics for systems of $n$ particles described in configuration space. Writing the Schr\"{o}dinger wave function as $\psi=Re^{iS/\hbar}$,
the motion of the $i$-th particle is given by
  \be 
    \tfrac{1}{m_i}{\bf\nabla}_i S({\bf x}_1,\ldots,{\bf x}_n) \ .
  \label{beable1}
  \ee
Also, insofar as one may assume that measurement results are recorded in some object's position (a pointer, ink on paper, ultimately somewhere in the brain) it follows that ordinary measurements of position are faithful.\footnote{The (non-trivial and generally non-faithful) measurement theory for observables other than position was worked out only by Bohm (1952). Note that Einstein's (1953) published objection to Bohm's theory was precisely that even for {\em macroscopic systems}, it violates faithful measurability in the case of momentum. We shall just need position measurements in the following.} In this sense, pilot-wave theory is a theory that satisfies with respect to position both the assumption of realism (in that it always has a value and a well-defined dynamics --- in this case deterministic) and that of faithful measurability. 

It is clear, at least qualitatively, that the 
theory predicts both interference and diffraction phenomena.
Indeed, around the nodes of the 
modulus $R$ of the wave function, the phase $S$ will behave very 
irregularly, so one expects that the particles
will be driven away from regions of configuration space where $R$ is
small. If one assumes additionally that particle positions are initially distributed according to $R^2$, one can show
that this form of the distribution is preserved over time (as de~Broglie himself explicitly remarks), and one obtains the same quantitative predictions as quantum mechanics.
As a matter of fact, de~Broglie predicted electron 
diffraction based on similar considerations well before 1927 when the detailed 
theory was worked out.\footnote{For a fuller historical summary, see Bacciagaluppi and Valentini (2009, esp.\ Chap.~2).} 

Now take the simple case of a plane wave travelling towards a screen with a single slit. The particle has a uniform velocity
perpendicular to the screen. If the initial position of the particle corresponds to the position of the slit, the particle will go through, otherwise it will not. We can imagine the screen being coated with photographic emulsion, so that the particle will be detected if and only if it does not pass through the slit. That is, in the case of a negative result, we have a measurement of position that is both faithful and apparently as non-invasive as can be. 

Nevertheless, {\em the subsequent dynamics of the particle is affected}. Indeed, after the particle has passed through the slit, the relevant pilot wave is no longer the incident plane wave, but only the spatially narrow component that has gone through the slit along with the particle. The rest of the wave has not `collapsed', but has interacted with the screen, and is no longer relevant to the subsequent motion of the particle (unless the different components are later brought to reinterfere together). This spatially narrow wave function, however, has large transverse momentum components, and the particle is diffracted accordingly. 
We thus have a straightforward violation of non-invasive measurability (as made precise by (\ref{3})), even though the theory satisfies both realism and faithful measurability with respect to position.\footnote{The failure of non-invasive measurability in pilot-wave theory is also pointed out by Kofler and Brukner (2013) and by Maroney and Timpson (in preparation).}

But now, the situation is quite analogous if we try to model the macroscopic case considered by Leggett and Garg. We have a two-valued quantity $Q$ that always has a value. The assumption that measurements of $Q$ are faithful implies (\ref{2}), and thus
  \begin{equation}
    \rho(Q^i_i)=\rho(Q_i) \ ,
  \label{beable2}
  \end{equation}
\ie at all times $t_i$ the distribution of values of $Q$ is the same as that for an ensemble of measurement results. If we now assume that the latter is correctly given by quantum mechanics, we obtain that the single-time distributions $\rho(Q_t)$ for the values of $Q$ at $t$ must be given by
  \be
    \rho(Q_t=\pm 1)=\bra{\psi(t)}P_{\pm 1}\ket{\psi(t)} \ ,
  \label{constraint}
  \ee
where $P_{\pm 1}$ are the eigenprojections of $Q$ corresponding to the eigenvalues $\pm 1$. Since $Q$ takes no other values, it tunnels discontinuously back and forth between the two potential wells. 

This can be easily modelled using a stochastic dynamics.\footnote{This is so whether or not one might think there is some further mechanism that determines also the times of the macroscopic tunnellings (which shall be of no concern to us).} Bell (1986) provides an explicit discussion of the most general Markovian dynamics for the values of a discrete observable (or `beable' in his terminology) obeying the constraint (\ref{constraint}), as follows.\footnote{For a more detailed discussion, see \eg Bacciagaluppi and Dickson (1999), Bub (1999, Chap.~5), or Vink (1993). Note that under certain circumstances, one can then indeed prove that ideal measurements of the beable are faithful; see the remarks in Bacciagaluppi and Dickson (1999, Section 5.2) and references therein.} 

Any Markovian dynamics can be reconstructed from its infinitesimal transition probabilities, in this case 
  \be
    t_{ji}(t):=\lim_{\epsilon\ra 0}\frac{\rho(Q_{t+\epsilon}=j|Q_t=i)}{\epsilon} \ .
  \label{eq:infinitesimal}
  \ee
The classic results from the 1930s and 1940s needed to show this do not cover the case in which the single-time probabilities have zeros, which happens periodically for (\ref{constraint}) under the evolution (\ref{a}); but Georgii and Tumulka (2005) have provided the required generalisations. The infinitesimal transition probabilities can in turn be constructed by solving the following master equation
  \begin{equation}
  \label{eq:master}
    \dot{\rho}_j(t) = \sum_i \Big( t_{ji}(t)\rho_i(t) - t_{ij}(t)\rho_j(t) \Big)
  \end{equation}
(which follows from the law of total probability). This is in fact a continuity equation for the probability, if one identifies the probability current $j_{ji}(t)$ as
  \be
    j_{ji}(t)=t_{ji}(t)\rho_i(t) - t_{ij}(t)\rho_j(t) \ .
  \label{eq:sufficient}
  \ee

If we insert into (\ref{eq:sufficient}) the quantum mechanical probability current,
  \be
  \label{eq:current2}
    j_{ji}(t) = 2 \Im 
    \bra{\psi(t)}P_jHP_i\ket{\psi(t)} \ ,
  \ee
we obtain a solution to (\ref{eq:master}), and we can explicitly construct the desired infinitesimal transition probabilities, for instance as
  \begin{equation}
    t_{ji}(t):=\max\Big\{0,\frac{j_{ji}(t)}{\rho_i(t)}\Big\} \ .   
  \label{eq:bell}
  \end{equation}
This is Bell's choice, and it yields a very precise discrete Markovian analogue of de~Broglie and Bohm's pilot-wave theory. Indeed, Vink (1993) has shown that one recovers de~Broglie--Bohm theory as the appropriate continuum limit of such a theory, when the beable is a discretisation of position. 

We thus see that the assumptions of macroscopic realism and faithful measurability for the observable $Q$, together with the constraint that the values of $Q$ be distributed according to the usual Born probabilities, naturally leads to a pilot-wave dynamics for $Q$. On the other hand, as we pointed out in the example of de~Broglie--Bohm theory, pilot-wave theories do not satisfy non-invasive measurability: a measurement of the beable of the theory, even a negative-result one, will quite generally change the effective pilot-wave of the system, and thus affect its subsequent dynamics. This shows that, even though Leggett and Garg's arguments may establish the plausibility of faithful measurability given macroscopic realism, their further assumption of non-invasive measurability is unwarranted. Thus, even assuming macroscopic realism, one need not expect the Leggett--Garg inequality to be violated.\footnote{In Section~\ref{analogy} we shall mention how in certain circumstances it is easy to calculate how the subsequent dynamics of $Q$ is affected, and thus to see explicitly how the invasivity of the measurements explains the violation of the inequality.} 

In the next two sections, we shall discuss how to think of such a violation. In any case, however, the above shows that the violation of the inequality does not provide a good argument against macroscopic realism as such.

\section{Contextuality and signalling}\label{context}
Bell-type inequalities have been recognised as providing necessary and sufficient conditions for the existence of joint distributions ever since the work of Suppes and Zanotti (1981) and of Fine (1982), with `contextuality' usually defined as the non-existence of such a joint distribution. This kind of ana\-lysis, however, standardly presupposes that certain relevant {\em pairs\/} of observables have joint distributions, and the question is whether the set as a whole has. Thus, in particular, for these pairs we have that the marginal distribution for the results of one measurement does not depend on having performed the other measurement. More generally, one requires that the probabilities for the results of one measurement do not depend on the inputs of the other measurement (`no-signalling' or `marginal selectivity', as the condition is more generally known outside of the literature on quantum mechanics). 

Indeed, if  for instance measuring $A$ alongside with $B$, or measuring $A$ on its own, yield different probability distributions for $A$, then $A$ does not actually refer to the same random variable in the two contexts, and one ought to distinguish two different random variables, say $A^B$ and $A$. No-signalling can of course be enforced by assuming spacelike separation of the measurements considered (or using the quantum mechanical no-signalling theorem), but this strategy is clearly precluded when considering sequential measurements as in the Leggett--Garg scenario. 

In such cases, it is less obvious what a violation of the corresponding Bell inequalities means, since we know already that the existence of joint probabilities is precluded.\footnote{In particular, it follows that the presence of signalling already rules out the conjunction of macroscopic realism and faithful and non-invasive measurement. This can also be seen directly, e.g.\ since
  \[
    \rho(Q^{ik}_k)=\rho(Q^i_k)=\sum_{Q^i_i=\pm1}\rho(Q^i_k|Q^i_i)\rho(Q^i_i)
    =\sum_{Q_i=\pm1}\rho(Q_k|Q_i)\rho(Q_i)=\rho(Q_k)=\rho(Q^k_k) \ ,
  \]
thus $\rho(Q^{ik}_k)=\rho(Q^k_k) 
$ for all $k>i$ (and in fact for all $k\neq i$, if one assumes that earlier transition probabilities are independent of whether one carries out a later measurement), as discussed also by Maroney and Timpson (in preparation) and in detail by Kofler and Brukner (2013).}

This question has been recently taken up and given a beautiful treatment by Dzhafarov and Kujala (2014a,b), who provide a principled criterion for distinguishing contextuality from mere violation of marginal selectivity. For the case of Leggett and Garg, they obtain, as the necessary and sufficient criterion for such contexuality, the violation of a modified Suppes--Zanotti inequality, namely  
  \begin{multline}
    -1-2\Delta_0\leq \exp{Q^{12}_1Q^{12}_2}+\exp{Q^{23}_2Q^{23}_3}+\exp{Q^{13}_1Q^{13}_3}\leq \\[1ex]
    \leq1+2\Delta_0+2\min\{\exp{Q^{12}_1Q^{12}_2},\exp{Q^{23}_2Q^{23}_3},\exp{Q^{13}_1Q^{13}_3}\}\ ,
  \label{SZmodified}
  \end{multline}
where 
  \be  
    \Delta_0:=\tfrac{1}{2}\Big(|\exp{Q^{12}_2}-\exp{Q^{23}_2}|+|\exp{Q^{13}_3}-\exp{Q^{23}_3}|\Big)  
    \label{context1}
  \ee
is the measure for the violation of marginal selectivity (i.e.\ the measure for signalling) obtained uniquely if one minimises the unobservable probabilities 
  \be
    \rho(Q^{12}_1-Q^{13}_1),\; \rho(Q^{12}_2-Q^{23}_2),\; \rho(Q^{13}_3-Q^{23}_3) 
  \ee
under the constraint that the expectation values 
  \be
     \exp{Q^{12}_1},\,\exp{Q^{13}_1},\,\exp{Q^{12}_2},\,\exp{Q^{23}_2},\,\exp{Q^{13}_3},\,\exp{Q^{23}_3} 
  \ee
are given.

Compared to (\ref{d'}), in order to evaluate (\ref{SZmodified}) we also need to calculate $\Delta_0$, but it is easy to show that for $i<j$,
  \be
    \exp{Q^{ij}_j}=\cos(\tfrac{\Delta E}{\hbar}(t_j-t_i))\cos(\tfrac{\Delta E}{\hbar}(t_i-t_0)) \ .
  \label{context2}
  \ee
Also, since for $i>j$ we have $\exp{Q^{ji}_j}= \exp{Q^{j}_j}$, and the latter is given by (\ref{b}), we  have 
  \be
    \exp{Q^{ji}_j}=\cos(\tfrac{\Delta E}{\hbar}(t_j-t_0)) \ .
  \label{context3}
  \ee
Thus,
  \begin{multline}
    \Delta_0=\tfrac{1}{2}\Big(|\cos(\tfrac{\Delta E}{\hbar}(t_2-t_1))\cos(\tfrac{\Delta E}{\hbar}(t_1-t_0))-
    \cos(\tfrac{\Delta E}{\hbar}  (t_2-t_0))|+  \\
    \qquad |\cos(\tfrac{\Delta E}{\hbar}(t_3-t_1))\cos(\tfrac{\Delta E}{\hbar}(t_1-t_0))-
    \cos(\tfrac{\Delta E}{\hbar}(t_3-t_2))\cos(\tfrac{\Delta E}{\hbar}(t_2-t_0))|\Big) \ .
  \label{context4}
  \end{multline}
Choosing again (\ref{h}), which maximally violates the original Leggett--Garg inequality, we obtain
  \be
    \Delta_0=\tfrac{1}{2}\Big(|-\tfrac{1}{2}\cos(\eta)-\cos(\eta+\tfrac{2\pi}{3})|+
    |-\tfrac{1}{2}\cos(\eta)+\tfrac{1}{2}\cos(\eta+\tfrac{2\pi}{3})|\Big) \ ,
  \label{Delta}
  \ee
where we have set $\eta:=\tfrac{\Delta E}{\hbar}(t_1-t_0)$. The range of (\ref{Delta}) is
  \be
    \tfrac{3}{8}\leq\Delta_0\leq\tfrac{3}{4} 
  \label{range}
  \ee
(as can easily be seen numerically), so that the tightest lower bound of the modified Leggett--Garg  inequality (\ref{SZmodified}) becomes $-1.75$, and the inequality is {\em always\/} satified.

This is perhaps surprising, but even more surprising is the following. 

As mentioned in Section~\ref{inequalities}, the product expectation value (\ref{c}) is independent of the initial state $\ket{\psi(0)}$ of the SQUID. Therefore, the amount by which the Leggett--Garg inequality (\ref{d'}) is violated is also independent of $\ket{\psi(0)}$, or even of whether the initial state is a pure state of the form (\ref{a}) or an arbitrary mixture of such states. By contrast, (\ref{context2}) and (\ref{context3}) do depend on the initial state, and thus so does the term $\Delta_0$ that enters the modified Suppes--Zanotti inequality (\ref{SZmodified}). Therefore --- unlike the case of the original Leggett--Garg inequality --- the amount by which the modified inequality may be violated (and thus the amount of contextuality in the sense of Dzhafarov and Kujala) depends in fact on the initial state of the SQUID.

Now take the case in which the initial state is an equal-weight mixture of $\ket{\psi_R}$ and $\ket{\psi_L}$, \ie the maximally mixed state. Obviously one has $\exp{Q^j_j}=0$ for all $j$, but it is also easy to check explicitly that  $\exp{Q^{ij}_j}=0$ for all $i<j$. Indeed, if the initial state is $\ket{\psi_R}$, we have 
\be
p_R(Q^{ij}_j=1)=\cos^2(\tfrac{\Delta E}{\hbar}(t_j-t_i))\cos^2(\tfrac{\Delta E}{\hbar}t_i)+\sin^2(\tfrac{\Delta E}{\hbar}(t_j-t_i))\sin^2(\tfrac{\Delta E}{\hbar}t_i) \ ,
\label{indeed1}
\ee
while if the initial state is $\ket{\psi_L}$, we have 
\be
p_L(Q^{ij}_j=1)=\cos^2(\tfrac{\Delta E}{\hbar}(t_j-t_i))\sin^2(\tfrac{\Delta E}{\hbar}t_i)+\sin^2(\tfrac{\Delta E}{\hbar}(t_j-t_i))\cos^2(\tfrac{\Delta E}{\hbar}t_i) \ .
\label{indeed2}
\ee
Thus,  
\be
\tfrac{1}{2}p_R(Q^{ij}_j=1)+\tfrac{1}{2}p_L(Q^{ij}_j=1)=\tfrac{1}{2} \ ,
\label{indeed3}
\ee
and the probabilities if the state is intially maximally mixed are the {\em same\/} as without the previous measurement at $t_i$.\footnote{An alternative way of seeing this is to note that the sequential measurement of two observables on a two-level system in the maximally mixed state can be realised by performing one measurement each on two entangled two-level systems in the singlet state. In this case, it is obvious that the no-signalling condition is satisfied. This trick allows one to intertranslate variously between `contextuality' and `non-locality' results (cf. Bacciagaluppi 2014, pp.~30--32 and fn 34).} Consequently, we have $\Delta_0=0$, i.e.\ {\em there is no longer any violation of marginal selectivity}, and the violation of the original Leggett--Garg inequality implies also the violation of  the modified Suppes--Zanotti inequality. In other words, in the case of the maximally mixed state the previous situation is reversed: the temporal signalling disappears, and the Dhzafarov--Kujala analysis of the violation of the original Leggett--Garg inequality yields indeed a verdict of contextuality.\footnote{The following point was suggested to me by a remark by Acacio de Barros (which I hope I have not misinterpreted). The Dhzafarov--Kujala analysis aims at analysing the violation of some Bell inequality, e.g.\ (\ref{SZmodified}), in the presence of direct influence of one measurement on the distribution of results of a subsequent measurement. It then makes sense to analyse separately the different scenarios in which one tests for such violations. The above analysis implicitly assumed the scenario labelled (i) in footnote 3, in which one measures the three expectation values $\exp{Q^{ij}_iQ^{ij}_j}$ separately, and thus with the same initial state (whether pure or mixed) in all three experiments. If one tests for violation of (\ref{SZmodified}) using scenario (ii), then one may want to calculate $\Delta_0$ using the given (pure or mixed) initial state to evaluate the terms $\exp{Q^{12}_2}$ and $\exp{Q^{13}_3}$ but using the maximally mixed state (that results from the initial measurement in the context $123$) to evaluate the
terms $\exp{Q^{23}_2}$ and $\exp{Q^{23}_3}$, which then both equal 0. In the case of an initial pure state it then follows that $\Delta_0=\tfrac{1}{2}|\cos(\eta)|$, which ranges between 0 and $\tfrac{1}{2}$, and whether (\ref{SZmodified}) is violated or not becomes dependent on the value of $\eta$, i.e.\ that of $t_0$. It seems thus that although scenarios (i) and (ii) are equivalent for the purpose of testing the original Leggett--Garg inequality (\ref{d'}), they are inequivalent for the purpose of testing the modified inequality (\ref{SZmodified}). This point may require further analysis.}

\section{An analogy, and Schr\"{o}dinger's SQUID}\label{analogy}
The analysis of the last section has left us with a somewhat puzzling situation: the violation of the Leggett--Garg inequality --- which is independent of the initial state of the SQUID --- can be absorbed entirely into the violation of marginal selectivity (i.e.\ into the amount of temporal signalling), or interpreted entirely in terms of contextuality, depending on whether the initial state of the SQUID is pure or is the maximally mixed state. 

This situation, however, has a well-understood analogue in pilot-wave theory, which we shall describe in this final section. In pilot-wave theory, the violation of the Bell inequalities in an EPR scenario is explained as an effect of the non-local dynamics. For each EPR pair, the details of how the measurement on one side is carried out will determine the trajectory of the particle on the other side, and this non-locality of the dynamics enforces the correlations necessary for the violation of the Bell inequalities. It also allows for the possibility of signalling, if one imagines the initial particle positions as given.\footnote{This is just the very well-known fact that as a hidden variables theory pilot-wave theory exhibits {\em parameter dependence}, i.e.\ probabilistic dependence of the distant outcome in an EPR experiment on the parameters of the nearby measurement, and in this way can be used in principle (i.e.\ if one could know the hidden variables) to signal across an EPR set-up. See Jarrett (1984) and Shimony (1986) for the (closely related but not identical) classic analyses of non-locality along these lines.} However, if the initial positions are unknown (more precisely if one assumes the positions are distributed according to the usual quantum mechanical measure), then the possibility of signalling is washed out. 

To make the analogy clearer, let us spell out the sense in which results of measurements in pilot-wave theory are determined by the context of measurement.\footnote{One needs to distinguish this notion of measurement contextuality from the probabilistic notion of contextuality of the last section. Indeed, as we have seen, the measurement contextuality of pilot-wave theory is related to the possibility of signalling, while according to Dhzafarov and Kujala probabilistic contextuality is defined {\em over and above\/} the amount of signalling present.} 

(a) Local contextuality: the result of a spin measurement on Alice's side of the EPR set-up is in general jointly 
determined by the position of the particle on her side before her measurement and by the details of the experimental 
arrangement on her side.  (The same is true on Bob's side.)\footnote{In the EPR--Bohm spin scenario, the relevant details are the  polarity and gradient of the magnetic field. For a clear and accessible discussion, see e.g.\ Barrett (1999, Section~5.2).}

(b) Non-local contextuality: if before Alice's measurement Bob
has performed a measurement of spin, then in general the result of Alice's measurement depends also on the position of Bob's particle and the experimental details on his side. In the special case in which Alice measures spin in the same direction as Bob, then not only do the initial position of the electron on Bob's side and the details of his experimental arrangement jointly determine his own measurement result, but the non-locality of the dynamics now forces the electron on Alice's side to exhibit the opposite spin, {\em regardless\/} of its initial position and of the details of Alice's experimental arrangement. 

Now assume an ensemble in which the initial positions of the two electrons are given (`pure hidden state'). Then, in general, the final position distribution for Alice's electron will depend on whether or not Bob has performed the measurement on his side, i.e.\ we have a violation of marginal selectivity. 

Assume instead that we have an appropriately mixed hidden state, specifically the usual quantum distribution of initial positions. If the initial position distribution is the usual quantum mechanical one, then so is the final position distribution, but that means the detections on Alice's screen are distributed just as quantum mechanics predicts. And we know from the no-signalling theorem that it makes {\em no difference\/} to this distribution whether or not Bob has performed a previous measurement on his side. Thus, if we assume an appropriately mixed hidden state, marginal selectivity is restored: the `noise' induced by the distribution of positions on Bob's side obliterates the `signal' he could in principle send by choosing to perform his measurement. 
 
We can make the analogy even closer by dispensing with the EPR scenario, and looking instead at a Bell-type pilot-wave model of the Leggett--Garg scenario as in Section~\ref{invasive}, with $Q$ as beable. In this case,\footnote{Using the techniques described in Bacciagaluppi and Dickson (1999, Section 5.2) and references therein.} one can show that at least under certain circumstances the probabilities for the values of the beable $Q$ conditional on the initial values $Q=\pm1$ are the same as the quantum mechanical probabilities for $Q$ conditional on the initial state being $\ket{\psi_R}$ or $\ket{\psi_L}$, for which we have seen in Section~\ref{context} that we have signalling in excess of the violation of the Leggett--Garg inequality. Thus, assuming a beable theory as in Section~\ref{invasive}, even though there is no signalling if the quantum state is maximally mixed, we have signalling in principle if the beables are known.\footnote{Note the (at least partial) analogy to how in standard pilot-wave theory the violation of the Bell inequality in the EPR case is explained through the possibility of signalling in principle, although at the quantum level there is no signalling. It will be interesting to explore the analogy further, making explicit the parallels between the Dzhafarov--Kulaja analysis of contextuality and the Jarrett--Shimony analysis of non-locality.}

Returning to the purely quantum mechanical case, we can now say the following. If the initial state is either definitely the pure state 
$\ket{\psi_R}$ or definitely the pure state $\ket{\psi_L}$, then we can affect the distribution of $Q$ at $t_j$ by performing a measurement of $Q$ at an earlier time $t_i$. If the initial state is an equal weight mixture of the two states, and we only know with probability $\tfrac{1}{2}$ that we have $\ket{\psi_R}$ or $\ket{\psi_L}$, then our performing a measurement at $t_i$ will produce one half of the time the distribution for $Q$ at $t_j$ given by (\ref{indeed1}), and the other half of the time the distribution given by (\ref{indeed2}), without us being able to control which. But as we have seen, the resulting averaged distribution (\ref{indeed3}) is the uniform distribution, which is exactly the same as predicted by the initial equal-weight mixture for the case in which we do not perform any measurement. Thus, marginal selectivity is restored on average. 

This, however, cannot be the full explanation of the violation of the Leggett--Garg inequality, because in quantum mechanics a mixed state need not be ignorance-interpretable. We could entangle the SQUID with a microscopic probe, e.g.\ as described by Leggett and Garg themselves (1985, p.~859), so that  there is indeed no matter of fact whether it is in the state $\ket{\psi_R}$ or $\ket{\psi_L}$. Or one could imagine a sealed box with an atom with half-life of one hour and a mechanism that prepares the SQUID in the state $\ket{\psi_R}$ or $\ket{\psi_L}$ depending on whether at the end of the hour the atom has decayed or not. In the case of such `Schr\"{o}dinger SQUIDs',  it thus seems there is no way out of interpreting the violation of the Leggett--Garg inequality as an indication of genuine contextuality.

{\small

\section*{\normalsize Acknowledgements}
This paper was prompted by talks of and discussions with Ehtibar Dzhafarov and Acacio de Barros at the conference on `Quantum Theory: from Problems to Advances' at Linnaeus University, V\"{a}xj\"{o}, Sweden, 9--12 June 2014. Many thanks to Ehti and especially to Acacio for detailed subsequent correspondence and comments on previous drafts of this paper, and to Ehti for the opportunity to present this material at the Weiner Memorial Lectures at Purdue University 1--3 November 2014. Many thanks also to Johannes Kofler for alerting me to his beautiful paper with \v{C}aslav Brukner, to Chris Timpson for sharing a draft of his paper with Owen Maroney, and to an anonymous referee for comments.

\

\section*{\normalsize References}
\noindent Bacciagaluppi, G. (2014),`Quantum Probability: An Introduction',\\ \url{http://philsci-archive.pitt.edu/10614/}. Slightly abridged version to appear in A.~H\'{a}jek and C.~Hitchcock (eds.), {\em The Oxford Handbook of Probability and Philosophy} (Oxford: Oxford University Press).

\noindent Bacciagaluppi, G., and Dickson, M. (1999), `Dynamics for Modal Interpretations', {\em Foundations of Physics\/} {\bf 29}, 1165--1201.

\noindent Bacciagaluppi, G., and Valentini, A. (2009), {\em Quantum Theory at the Crossroads: Reconsidering the 1927 Solvay Conference} (Cambridge: Cambridge University Press).

\noindent Barrett. J. A. (1999), {\em The Quantum Mechanics of Minds and Worlds} (Oxford: Oxford University Press).

\noindent Bell, J.~S. (1986), `Beables for Quantum Field Theory',  {\em Physics Reports\/} {\bf 137}, 49--54. Reprinted in {\em Speakable and Unspeakable in Quantum Mechanics} (Cambridge: Cambridge University Press,  1987), pp.~173--180.

\noindent Bohm, D. (1952), `A Suggested Interpretation of the Quantum Theory in Terms of ``Hidden''
Variables, I and II', {\em Physical Review\/} {\bf 85}, 166--179 and 180--193.

\noindent de Broglie, L. (1928), `La nouvelle dynamique des quanta', in H. A. Lorentz (ed.), {\em \'{E}lectrons et Photons: Rapports et Discussions du Cinqui\`{e}me Conseil de Physique Solvay} (Paris: Gauthiers-Villars), pp.~105-132. Translated as `The New Dynamics of Quanta', in Bacciagaluppi and Valentini (2009), pp.~341--363.    

\noindent Bub, J. (1999), {\em Interpreting the Quantum World} (Cambridge: Cambridge University Press, paperback edition with corrections).

\noindent Chakravarty, S., and Leggett, A. J. (1984), `Dynamics of the Two-State System with Ohmic Dissipation',  {\em Physical Review Letters\/} {\bf 52}, 5--8.

\noindent Dzhafarov, E., and Kujala, J. V. (2014a), `Probabilistic Contextuality in EPR/Bohm-type Systems with Signaling Allowed', \url{http://arxiv.org/pdf/1406.0243v3.pdf}.

\noindent Dzhafarov, E., and Kujala, J. V. (2014b), `Generalizing Bell-type and Leggett--Garg-type Inequalities to Systems with Signaling', \url{http://arxiv.org/pdf/1407.2886v2.pdf}.

\noindent Einstein, A. (1953),  `Elementare \"{U}berlegungen zur Interpretation der Grundlagen der Quantenmechanik', in {\em Scientific Papers Presented to Max Born on his Retirement from the Tait Chair of Natural Philosophy in the University of Edinburgh\/} (Edinburgh: Oliver and Boyd), pp.~33--40.

\noindent Fine, A. (1982), `Hidden Variables, Joint Probability, and the Bell Inequalities', {\em Physical Review Letters\/} {\bf 48}, 291--295.

\noindent Georgii, H.-O., and Tumulka, R. (2005), `Global Existence of Bell's Time-In\-ho\-mo\-ge\-neous Jump Process for Lattice Quantum Field Theory', {\em Markov Processes and Related Fields\/} {\bf 11}, 1--18, \url{http://arxiv.org/pdf/math/0312294v2.pdf}.

\noindent Jarrett, J.~P. (1984), `On the Physical Significance of the Locality Conditions in the Bell Arguments', {\em No\^{u}s\/} {\bf 18}, 569--589. 

\noindent Kofler, J., and Brukner, \v{C}. (2013), `Condition for Macroscopic Realism beyond the Leggett--Garg Inequalities', {\em Physical Review\/} {\bf A 87}(5), 052115, \url{http://arxiv.org/abs/1207.3666}.

\noindent Leggett, A. (2002), `Testing the Limits of Quantum Mechanics: Motivation, State
of Play, Prospect', {\em Journal of Physics: Condensed Matter\/} {\bf 14}, R415--R451.

\noindent Leggett, A.~J., and Garg, A. (1985), `Quantum Mechanics versus Macroscopic Realism: Is the Flux There when Nobody Looks?', {\em Physical Review Letters\/} {\bf 54}, 857--860.

\noindent Maroney, O., and Timpson, C., (in preparation), `Quantum- vs Macro-Realism: What does the Leggett--Garg Inequality Actually Test?'.

\noindent Schr\"{o}dinger, E. (1934), `\"{U}ber die Unandwenbarkeit der Geometrie im Kleinen', {\em Die Naturwissenschaften\/} {\bf 22}, 518--520.

\noindent Shimony, A. (1986), `Events and Processes in the Quantum World', in R. Penrose and C. Isham (eds.), {\em Quantum Concepts in Space and Time\/} (Oxford: Clarendon Press), pp.~182--203. Repr.\ in A. Shimony, {\em Search for a Naturalistic World View}, Vol. II (Cambridge: Cambridge University Press, 1993), pp.~140--162.

\noindent Suppes, P., and Zanotti, M. (1981), `When are Probabilistic Explanations Possible?', {\em Synthese\/} {\bf 48}, 191--199.

\noindent Vink, J. (1993), `Quantum Mechanics in Terms of Discrete Beables', {\em Physical Review\/} {\bf A 48}, 1808--1818.

}

\end{document}